\documentclass[floatfix,preprintnumbers,superscriptaddress,showpacs,prl,
               preprint]{revtex4}

\usepackage{graphicx}
\usepackage{amsmath,amsfonts,amssymb,bm}

\begin{document}

\bibliographystyle{apsrev}

\title{ Strong covalent bonding between two graphene layers }

\author{P.L. de Andres}
\affiliation{
Instituto de Ciencia de Materiales de Madrid (CSIC), Cantoblanco,
28049 Madrid, Spain} 

\author{R. Ram\'{\i}rez}
\affiliation{
Instituto de Ciencia de Materiales de Madrid (CSIC), Cantoblanco,
28049 Madrid, Spain} 

\author{J.A. Verg\'es}
\affiliation{
Instituto de Ciencia de Materiales de Madrid (CSIC), Cantoblanco,
28049 Madrid, Spain} 

\date{\today}

\begin{abstract}
We show that two graphene
layers stacked directly on top of each other
(AA stacking) form strong chemical bonds when the distance between planes is
0.156 nm. Simultaneously, C-C in-plane bonds are considerably
weakened from partial double-bond (0.141 nm) to single bond (0.154 nm).
This polymorphic form of graphene bilayer is
meta-stable w.r.t. the one bound by van der Waals forces at a
larger separation (0.335 nm) with an activation energy of 0.16 eV/cell.
Similarly to the structure found in hexaprismane,
C forms four single bonds in a geometry
mixing $90^{0}$ and $120^{0}$ angles. 
Intermediate separations between layers can be stabilized under external
anisotropic stresses showing a rich electronic structure changing
from semimetal at van der Waals distance, to metal when compressed, to
wide gap semiconductor at the meta-stable minimum.
\end{abstract}


\maketitle

Carbon shows one of the richest chemistry in the periodic table and
it is often found in allotropic forms. In molecules it is the basis 
for organic compounds, being central to different fields from biology to 
electronics in new materials. In solid state it shows very different
properties drifting from a soft metal (graphite, the most stable
phase at P=0 GPa, T= 0 K) to a hard wide gap 
semiconductor (diamond). New forms like fullerenes and nano-tubes have 
raised even more the interest in carbon for their potential applications. 
Recently, the realization of two-dimensional periodic systems made by the 
stacking of few graphene layers (FGL), going down to the single layer, 
has attracted much interest as the basis for new electronic 
devices\cite{novoselov04}. 
The peculiar linear dispersion found in the electronic band structure near 
the charge neutrality point (Dirac Point), where the carriers behave like 
mass-less chiral relativistic particles, translates in all sort of new 
phenomena related to transport properties on these 
systems\cite{heersche07}. 
Moreover, a variety of preparation techniques have been used
giving rise to samples showing
important differences\cite{rokuta99,horiuchi03,novoselov04,berger04,meyer07};
most notably: charge accumulation regions associated with physical
corrugation found in free standing graphene\cite{meyer07},
new properties induced in the graphene layers by
the epitaxial growth on a SiC substrate\cite{berger04}, 
or a modification of the stacking, from Bernal AB phase
to AA, found in carbon nanofilms grown from graphite
oxide\cite{horiuchi03}.
Accurate and detailed information on these samples 
obtained from structural techniques is sometimes 
difficult to interpret;
state-of-the-art theoretical total-energy methods are necessary
to understand the precise atomic and electronic structure of these 
films. In this paper, we report on the
formation of strong covalent bonds between graphene layers
stacked directly on top of each other (AA) at a distance
that is much smaller than $\sim 0.335$ nm, that is, the typical
distance for an alternating (AB) stacking based on weak
van der Waals forces (Fig. 1).
On this meta-stable polymorphic form of a graphene bi-layer
each carbon is bonded to the four 
nearest neighbours, at 0.154 and 0.156 nm for in-plane and 
out-of-plane bonds respectively. 
Under these conditions, the bi-layer is a wide gap
semiconductor (indirect gap of 0.91 eV). 
As a function of the separation between layers, transport properties of the 
un-doped AA stacking are rich: 
at large distances between planes (e.g., as found
in graphite) the system is very close to a semi-metal,
mostly dominated by the single graphene layer
properties. As the distance between layers decreases it is possible
to find interlayer distances and/or different 2D unit cell sizes where 
the bi-layer becomes metallic.

\begin{figure}
\includegraphics[clip,width=0.9\columnwidth]{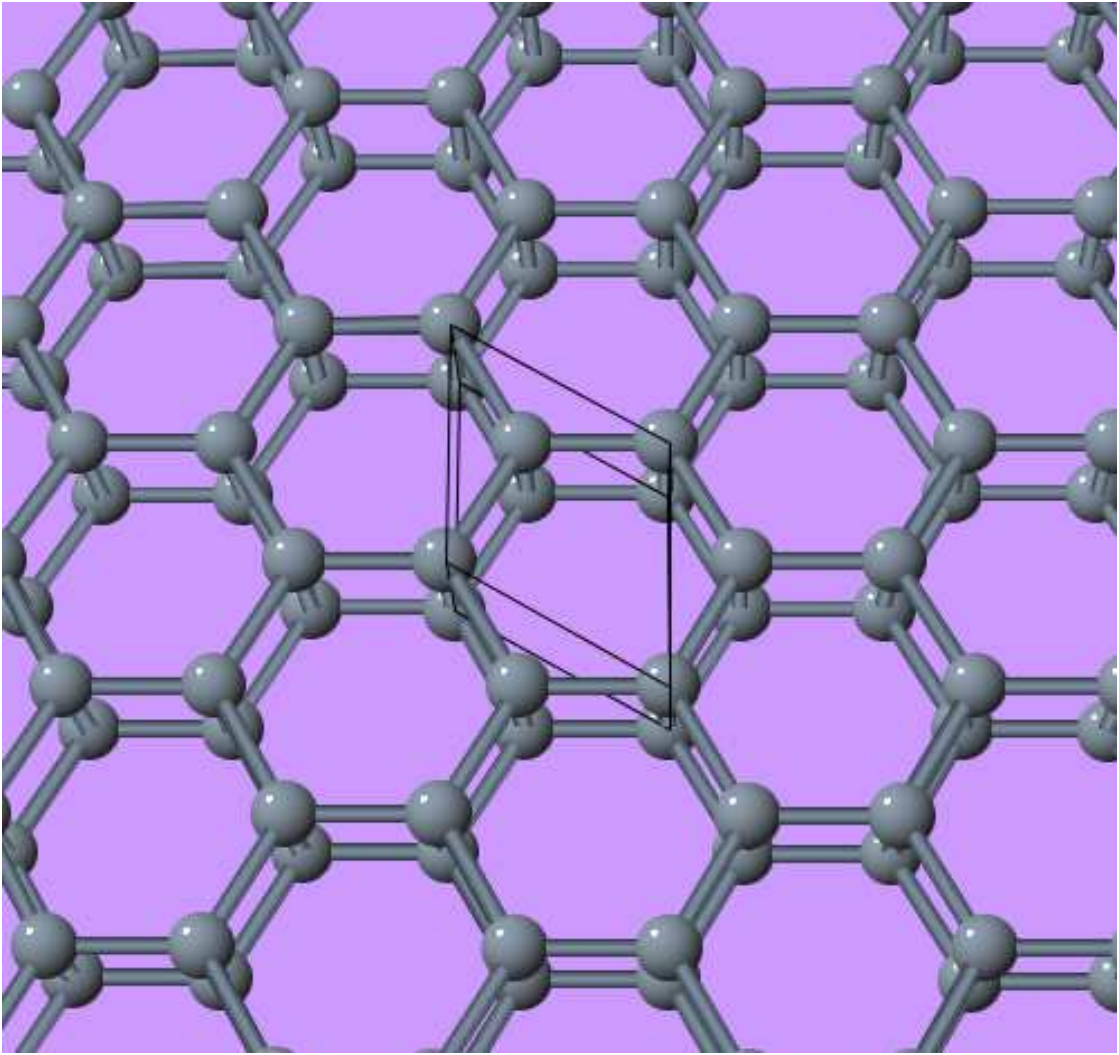}
\caption{(Color online) 
Meta-stable extended 2D carbon allotrope formed by two graphene layers 
at covalent C-C bond distance and direct on-top stacking (AA). 
The 2D unit cell is shown ($a=b=0.267$ nm, $\gamma=120^{o}$). 
}
\label{fig1}
\end{figure}

Our ab-initio calculations are based in
Density Functional Theory
(DFT)\cite{hohenberg65} 
and a local approximation
to describe exchange and correlation (LDA)\cite{kohn65,gga}. 
LDA calculations performed with CASTEP\cite{payne02,accelrys,precision} 
reproduce very well distances and angles for the strong
sp$^{2}$ bonds inside the graphene layer, 
predicting as the most stable configuration a honeycomb lattice 
with a C-C distance of 0.141 nm, 
and a bond population of 1.53.
Experimental value is 0.142 nm 
(fractional error less than 1{\%}), 
in between a carbon double bond 
(typical length 0.133 nm) and a single one (0.154 nm).  
A negligible charge transfer (0.3{\%}) takes place from 2s to 2p orbitals. 
These geometrical results, together with those obtained for electronic 
and vibrational properties, demonstrate the ability of DFT to describe 
the C-C bond at typical distances allowing
the formation of covalent bonds. 
A rough electron-counting picture for 
the graphene layer would be 
each C atom sharing one electron with each of the in-plane 
three nearest C neighbours, while the 
fourth electron is delocalized among them, 
making three stronger C-C bonds 
with a character somewhere in between a single and a double bond. 
In bulk graphite this fourth electron would be responsible for the 
appearance of pockets near the Fermi energy and the in-plane 
conductivity. 
This scenario makes plausible to use this extra electron to establish 
single bonds between carbons across the layers. 
While van der Waals interaction is weak and
not accurately described by a local DFT,
the formation of the new allotropic form of graphene 
bi-layer rather involves interactions between carbons
at shorter covalent bonding distances
where the DFT formalism is accurate and realistic.
This is independent of the basis choosen to solve
the equations;
we have checked that quantum chemistry calculations made
with localized basis sets and mixed functional methods
in small clusters concur with the
ones derived from plane-waves basis for extended 2D systems.

\begin{figure}
\includegraphics[clip,width=0.85\columnwidth]{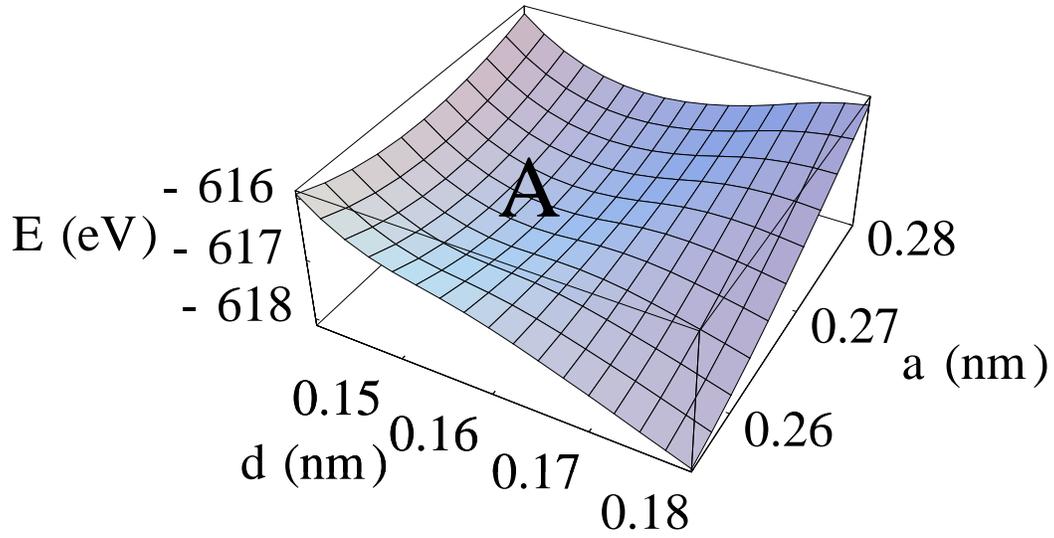}
\includegraphics[clip,width=0.85\columnwidth]{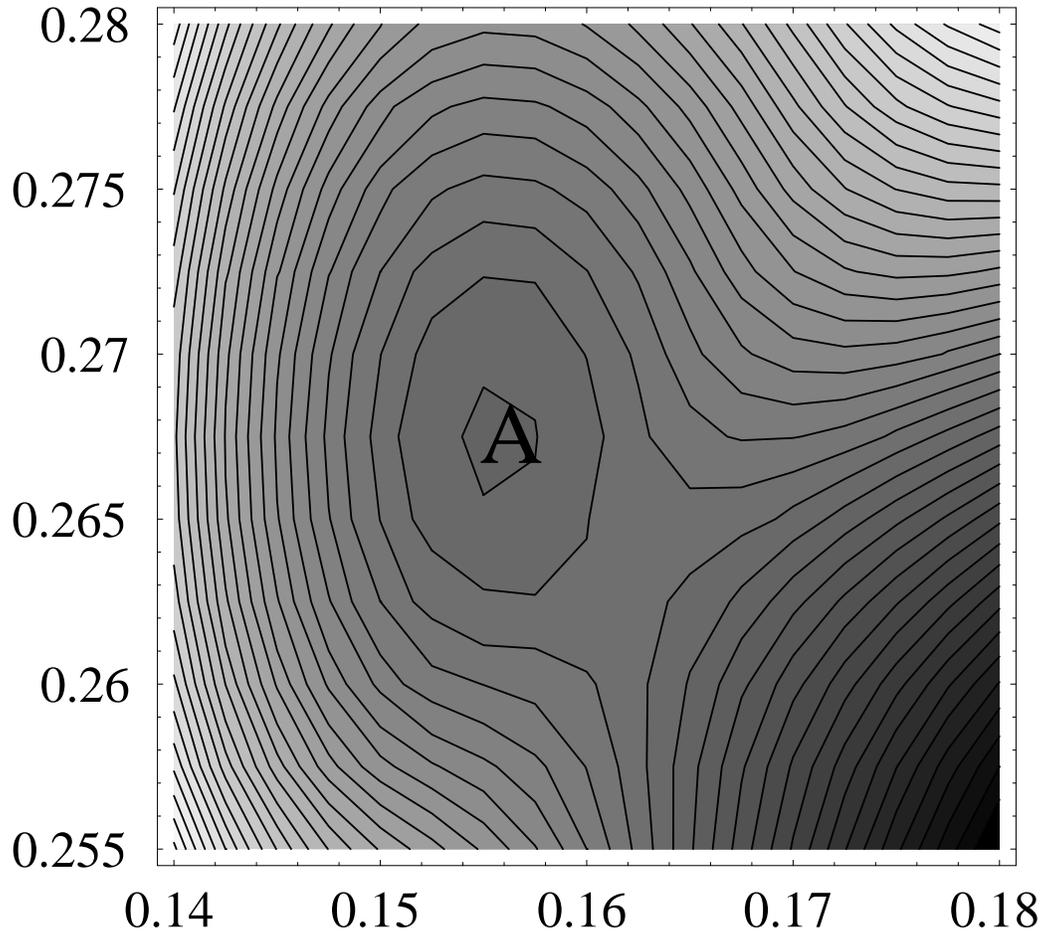}
\caption{(Color online)
First-principles total energy landscape
for the graphene bi-layer as a function of the 
distance between layers ($d$) and the 2D lattice parameter ($a$). 
Label A shows the local minimum reported
in this work at $d=0.156$ nm and $a=0.267$ nm 
corresponding to the formation of covalent bonds across layers.  
Contour lines start at -618 eV and go up in steps of 0.05 eV 
(the minimum in A is at -617.208 eV).
}
\label{fig2}
\end{figure}

Recent papers have investigated the electronic structure of
the standard alternating AB stacking since it is energetically
favoured over the AA stacking\cite{latil06}.
However, the expected energy difference is necessarily small due to the 
weak interaction between layers, about 0.02 eV/cell in our calculations.
The barrier to transform one stacking into the
other might be higher, but of the same order.
Therefore, we have investigated the AA bi-layer, searching for 
new structural configurations; we find a meta-stable energy minimum 
at about half the usual distance between layers in graphite.
This new structure implies an important lateral relaxation of the 
2D unit cell, and displays electronic 
properties quite different from the global van der Waals-like minimum. 
Fig. 2 shows a 2D total energy map for the system near the new minimum:
the meta-stable configuration appears around an interlayer distance 
$d=0.156$ nm and lattice parameter $a=0.267$ nm (label A in Fig. 2). 
The alternating Bernal stacking (AB) does not show a similar 
meta-stable local minimum in our calculations.
The reason for the different behaviour lies on the different
coordination of the C atoms in the bi-layer stacking AA and AB. 
While {\em all} C atoms can form an interlayer covalent
bond in the AA stacking, only half of the atoms have this
possibility for the AB case. Consequently, when a small separation is
forced in the AB bi-layer, buckling of both planes
can release stress efficiently, and
sp$^{3}$ coordination with nearly tetrahedral angles appears
(the resulting structure is a 2D diamond precursor).
In the AA case, the formation of a meta-stable configuration
is favoured because symmetry does not allow the relaxation of
structural strain by buckling. 
A similar idea has been put forward
to explain the meta-stability of n-prismanes\cite{jenkins00}.
Stacking of carbon layers with covalent
bonds accross layers seems unnoticed;
because this configuration is meta-stable it should require
contributing some external energy to the system. 
A natural way of doing this is to grow the layers
epitaxially on a substrate imposing a stretched length for the
2D unit cell. 
However, we should mention that the AA stacking has been reported
in the literature for some related system\cite{horiuchi03}. 

\begin{figure}
\includegraphics[clip,width=0.8\columnwidth]{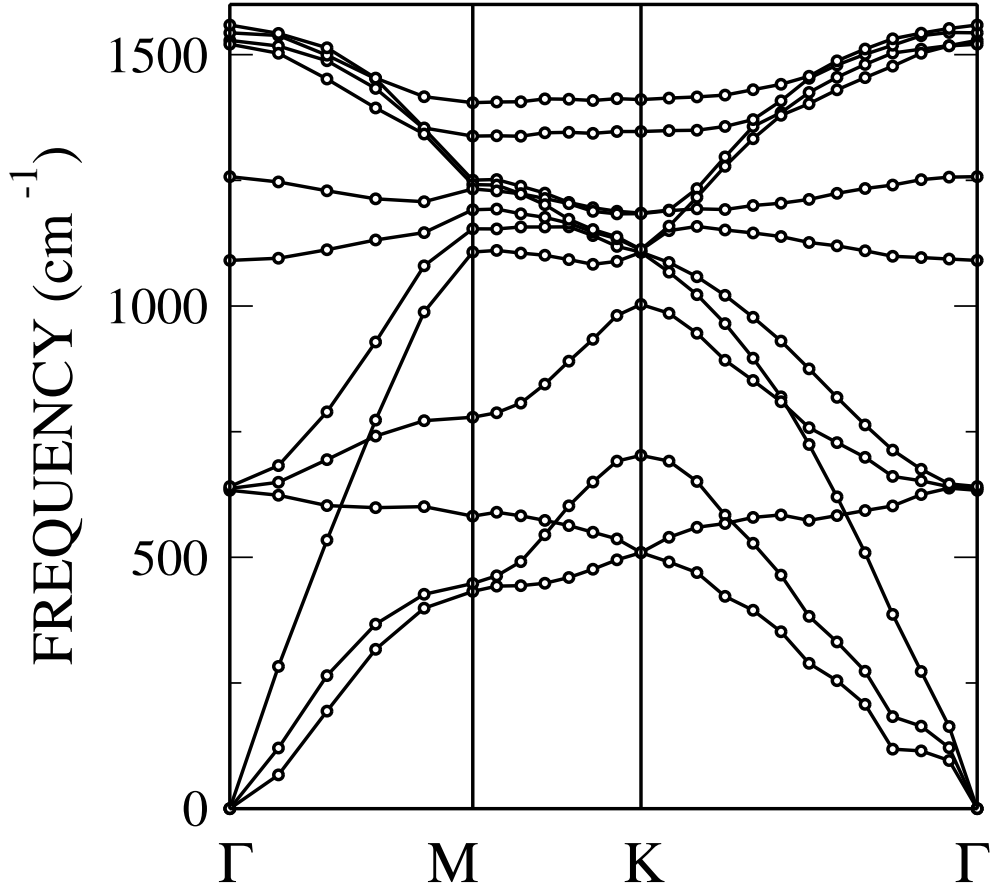}
\caption{Phonon spectrum
calculated at the local minimum A in 
Fig. 2. 
The x-axis samples the boundary of the irreducible 2D Brillouin zone. 
Lines between points are only meant to guide the eye.
}
\label{fig3}
\end{figure}

We have obtained the barrier to escape from minimum A to the global 
one (G) by applying first a Linear Synchronous Transit (LST) 
transition state search, followed by a Quadratic Synchronous Transit 
(QST) method. We find a barrier of $0.16 \pm 0.04$ eV/cell, 
allowing us to predict that the structure A is stable at room 
temperature. 
The energy barrier for the formation of the meta-stable state 
(from G to A) amounts to 4.80 eV/cell, the A configuration being 
4.64 eV/cell higher in energy than the G one. 
The path from A to the transition state involves a simultaneous 
modification of parameters, $d$ and $a$ (Fig. 2) 
due to the correlation between bonds 
formed in and out the planes. 
Boundary conditions keeping the parameter $a$ fixed to a given value 
make a different scenario with interesting consequences. 
If $a$ is kept at a constant value of $0.279$ nm, 
the local minimum A is established at $d=0.155$ nm 
and the barrier grows to 0.8 eV/cell. 
For a constant value of $a= 0.291$ nm, 
A becomes the global minimum, and the barrier from A to G 
goes to 1.7 eV/cell, 
A being lower in energy than G by 1.03 eV/cell. 

\begin{figure}
\includegraphics[clip,width=1.0\columnwidth]{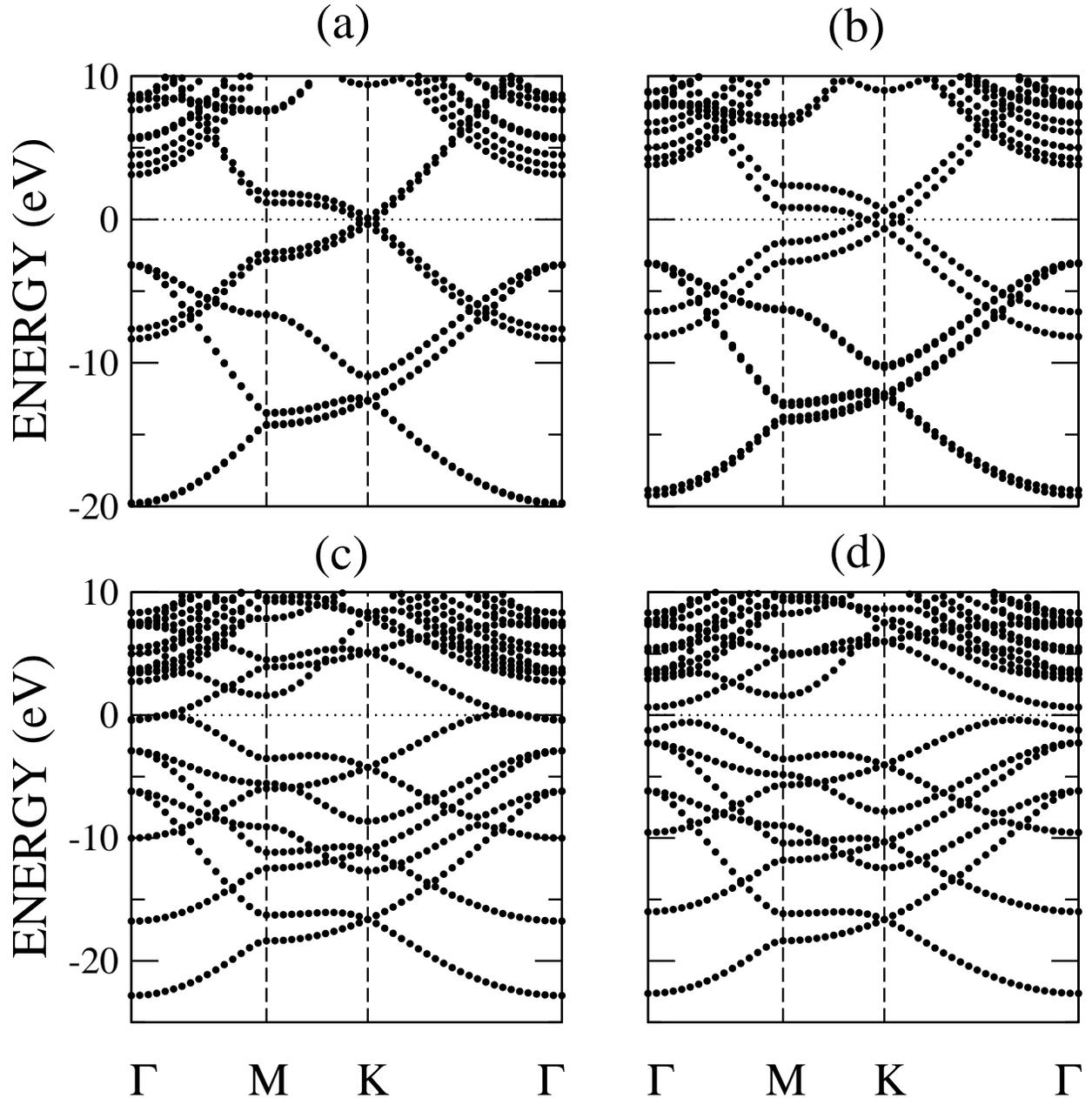}
\caption{
Evolution of the electronic band structure for the bi-layer 
as separation changes from van der Waals-like distance (a)
to the small separation allowing the chemical
bonding of graphene sheets (d). 
(a) $d=0.358$ nm, $a=0.243$ nm (global minimum G, semi-metallic);
(b) $d=0.300$ nm, $a=0.250$ nm (2D metal); 
(c) $d= 0.1625$ nm, $a= 0.2645$ nm (near the transition state, 2D metal); 
and 
(d) $d= 0.156$ nm, $a=0.267$ nm (local minimum A, insulator).  
Fermi energy is used as the origin for energies.
}
\label{fig4}
\end{figure}

Let us further characterize the new bonding configuration
after the formation of chemical bonds between C atoms
located in different layers.
The building of these bonds produces a weakening of the
sp$^{2}$-like in-plane bonds, that elongate from
0.141 nm to 0.154 nm.
In A, we observe a 0.1 electron charge transfer from the 
2p to the 2s orbital, and the formation of a single bond between 
carbons across the two graphene layers at 0.156 nm with a calculated 
bond order of 0.92. 
This distance is typical of single C-C bonding for 
substances like diamond, propane, etc.\cite{pauling}, 
supporting the formation of a chemical bond in place of the 
previous weak van der Waals interaction. 
We notice that similar strained carbon structures have been observed in 
molecular systems known as n-prismanes\cite{allinger83}. 
Quantum chemical calculations performed with the program
GAMESS\cite{gamess} confirm
the building of single C-C bonds across parallel carbon 
hexagonal rings saturated with H to form
the hexaprismane. C-C bond distances and angles
are similar to those found in the
graphene bi-layer (A).
In agreement with our periodic solid-state calculations, 
this is a meta-stable molecular configuration w.r.t.
van der Waals-like separation between two benzene molecules. 
Our calculations give a barrier between the meta-stable structure 
and the global minimum (C$_6$H$_6$-C$_6$H$_6$, one hexagonal ring)
of about 0.83 eV per C-C bond. 
This value decreases consistently as more rings are added; 
already for three hexagonal rings (C$_{13}$H$_{8}$-C$_{13}$H$_{8}$) 
it goes down to several tenths of eV per C-C bond. 
As the number of hexagonal rings increases this barrier 
converges to our result for the graphene bi-layer.

Fig. 3 gives the phonon spectra at the minimum A calculated 
with a linear response formalism\cite{refson06}. 
The phonon spectra has no dispersion in the direction perpendicular 
to C layers and shows that the new minimum is stable with 
respect to small displacements that preserve the unit cell 
area\cite{nota}. 
The optical branches around 1600 cm$^{-1}$ at $\Gamma$ 
can be compared with those measured for 
graphite\cite{maultzsch04}, 
although bonding in the layer is now weaker than for graphite. 
Near 1100 and 1250  cm$^{-1}$ we observe a couple of optical 
modes related to vibrations perpendicular to the layers that are 
similar in energy to that found for two
C$_{6}$H$_{6}$ rings (hexaprismane) vibrating against each other 
at C-C covalent distances. These may be used to experimentally
identify the bilayer.

Transport properties on FGL-based devices are determined by the 
band-structure of the material. Therefore, we study 
the electronic structure of the bi-layer for different structural parameters 
(size of the 2D unit cell, $a$, and separation between layers, $d$). 
A single graphene layer displays a semi-metallic character
with valence and conduction bands touching in 
the corners of the Brillouin zone,
\{{\bf K}\},
and the dispersion relation being linear.
At the van der Waals-like separation between layers (0.358 nm), 
the interaction is weak, but already a marginal 2D metal starts to 
form. The 2D Fermi circle is centred at the corners of the Brillouin 
zone, {\bf K}, with a very small radius and the density of states 
at the Fermi energy is nearly zero (Fig. 4a). 
We notice that in the AA stacking
the bands near {\bf K} are still linear, unlike
the AB stacking where the bands approach {\bf K} quadratically\cite{latil06}.
A new situation emerges if the two layers are forced to
get closer to each other.
Fig. 4b shows the band structure for such a
non-equilibrium configuration ($a=0.250$ nm, $d=0.300$ nm).
For this geometry, repulsive forces on atoms on each layer are 0.024 eV/nm. 
A comparison between panels a and b in Fig. 4 shows how the radius 
of the Fermi circle increases, yielding a distinctively non-zero 
density of states and making the bi-layer a 2D metal. 
This picture is still valid near the transition state, 
where the Fermi line is approaching the symmetry point $\Gamma$ 
in the Brillouin zone (Fig. 4c). 
Further down the distance between the two layers, the system develops 
strong single covalent bonds, and the bi-layer becomes a wide gap 
semiconductor.

Finally, we have explored the role of external stresses on the 
bi-layer by applying in-plane tensile stresses of 
$\sigma_{xx}$=$\sigma_{yy}$= 3, 6 and 9 GPa.
As expected, by forcing the 2D unit cell to extend, 
the minimum at A is stabilized and the barrier grows to 0.43, 
0.88 and 1.4 eV/cell respectively. 
The local minimum A changes so the 2D unit cell size grows 
from 0.267 nm to 0.273, 0.280 and 0.289 nm respectively, 
while the two layers come closer together by a small distance 
(0.0008 nm for 6 GPa). 
We notice that around G the strain is approximately half the value 
around A (from 0.243 to 0.249 nm for the 6 GPa stress), 
a consequence of the existence of stronger sp$^{2}$ bonds. 

In conclusion, we have found a new polymorphic form for two extended 
flat 2D graphene layers stacked with AA sequence where carbon atoms 
located in atop positions establish new 
covalent bonds. This meta-stable configuration is not 
subject to thermodynamic instability
and shows a barrier 
large enough to make it feasible at room temperature.
As a function of the separation between the two layers, 
their electronic properties range from a 
semi-metal (layers far away apart) to a weak 2D metal 
(van der Waals distances, low density of states at the Fermi energy) 
to a stronger 2D metal (intermediate distances, higher density of 
states at the Fermi energy), and finally to a wide gap semiconductor 
(covalent bonding distance). External stresses can help to further 
stabilize these configurations, as well as to control the separation 
between layers. 
The new predicted semiconductor should allow
traditional doping with impurities (B, N) opening a well-defined
way towards strict 2D electronics.

This work has been financed by the CICYT (Spain) under contracts
MAT-2005-3866, MAT-2006-03741, FIS-2006-12117-C04-03, 
and NAN-2004-09183-C10-08. 
We acknowledge the use of the Spanish Supercomputing Network
and the CTI (CSIC).


\begin{thebibliography}{20}
\bibitem[{gga()}]{gga}
\bibinfo{note}{Including gradient corrections to improve the description of the
  exchange and correlation functional (GGA) does not qualitatively change our
  conclusions; it merely modifies the fine details (GGA overestimates bonding
  distances by a similar amount as LDA underestimates them, while the barrier
  to the global minimum slightly increases to 0.21 eV). We have decided to use
  LDA because using a simpler formalism yields a simpler physical
  interpretation.}
\bibitem[{pre()}]{precision}
\bibinfo{note}{Accuracy for total energies have been converged below 0.01 eV
  and for barriers below 0.001 eV by using energy cutoffs around 750 eV
  (norm-conserving pseudo-potentials) and a set of about 100 k-points sampling
  the Brillouin zone.}
\end{thebibliography}

\end{document}